# End-to-End Learning of Joint Geometric and Probabilistic Constellation Shaping

**Vahid Aref and Mathieu Chagnon**
*Nokia, Magirusstr. 8, 70469 Stuttgart, Germany*
*Firstname.lastname@nokia.com*

**Abstract:** We present a novel autoencoder-based learning of joint geometric and probabilistic constellation shaping for coded-modulation systems. It can maximize either the mutual information (for symbol-metric decoding) or the generalized mutual information (for bit-metric decoding).
© 2022 The Author(s)

## 1. Introduction

Deep learning has attracted great research interest for optical communication with various applications in system design, system identification and network monitoring [1]-[3]. For the physical layer, deep learning has shown a great potential either to optimize individual digital signal processing (DSP) blocks or, more interestingly, to optimize the whole DSP blocks as a cascade of neural networks (NNs), referred to as an autoencoder. To mention examples of the former approach, NN-based digital pre-distortion has been proposed to pre-compensate the nonlinear impairments of optical coherent transmitters [4,5], outperforming other pre-distortion techniques such as Volterra-series based solutions. At the receiver side, different NN-based equalizations have been proposed to mitigate optical fiber and transceiver nonlinearities [6-8]. Another example is the use of deep neural networks for digital back-propagation with optimized computational complexity [9]. In the end-to-end (E2E) learning approach, the transmitter, channel, and receiver of a communication system are implemented as an autoencoder, jointly trained to best match the outputs to the inputs and thus, to better communicate [10]. In optical fiber communications, the E2E learning has been applied for both intensity modulation with direct detection (IM/DD) systems [11] and coherent systems [12-15]. For the latter, E2E learning is more involved due to the complex interplay of nonlinearity and chromatic dispersion in optical fibers.

For coded-modulation systems, most of E2E learning studies focused so far on geometric constellation shaping, e.g. see [12-16]. Knowing how to optimize probabilistic constellation shaping (PCS) is of great interest as PCS can be implemented efficiently in hardware with binary forward error correction (FEC) codes. A challenge of optimizing PCS is how correctly estimate the gradients of the cost function in terms of the constellation points probabilities. It was proposed in [17] to sample from these probabilities using the so-called Gumbel-Softmax trick. As indicated in [18], this method is prone to numerical instabilities because of its approximations and sensitive extra hyper-parameters. The difficulty of sampling from the probabilities is bypassed in [18] by sampling uniformly from the constellation points. However, this technique becomes too complex for channel with memory like optical fiber. In this paper, we propose a new autoencoder-based learning which samples from the constellation probabilities with no approximation and it is applicable to many channels like optical fiber. We show that it can accurately optimize either the PCS alone, or both geometric and probabilistic shaping (GeoPCS) jointly, for either symbol-metric decoding (SMD), which maximizes the mutual information (MI), or bit-metric decoding (BMD), maximizing the generalized mutual information (GMI).

## 2. Learning of Geometric and Probabilistic Shaping

In a coded-modulation system, the input data is mapped to some constellation points and then transmitted over a channel. Due to design constraints like transmit power limitation, the information rate of the system depends on the geometry of the constellation points and their probability distribution. Let $\mathcal{C}_M = \{c_1, c_2, ..., c_M\}$ denote the set of constellation points with corresponding probabilities $\mathcal{P}_M = \{p_1, p_2, ..., p_M\}$. The objective is to maximize the system information rate by optimizing $\mathcal{C}_M$ and $\mathcal{P}_M$. Optimizing $\mathcal{C}_M$ is called geometric constellation shaping and optimizing $\mathcal{P}_M$ is called probabilistic constellation shaping. Our scheme can optimize either of them or both jointly.

The proposed autoencoder is depicted in Fig. 1. It includes a mapper (encoder) and two possible demappers (decoders). The mapper consists of three parts: (i) the sampler from trainable $\mathcal{P}_M$, (ii) the trainable $\mathcal{C}_M$ (iii) the symbol mapper. To generate a training symbols batch of size $B$, the sampler draws randomly $B$ indices from $\mathcal{P}_M$. One way of sampling is by taking each index $m$ about $Bp_m$ times, for $m = 1, ..., M$, and then randomly permuting all indices. This approach needs some careful integer rounding when $Bp_m < 1$. Then, each index $m$ maps to the normalized symbol $\tilde{c}_m = c_m / \sqrt{\sum_{i=1}^{M} p_i c_i^2}$ in the training batch. Note that the normalization is required to maintain the power constraint. One

---
[*] Names appear alphabetically; VA and MC jointly contributed to developing the learning algorithm. The implementation is mainly done by MC.

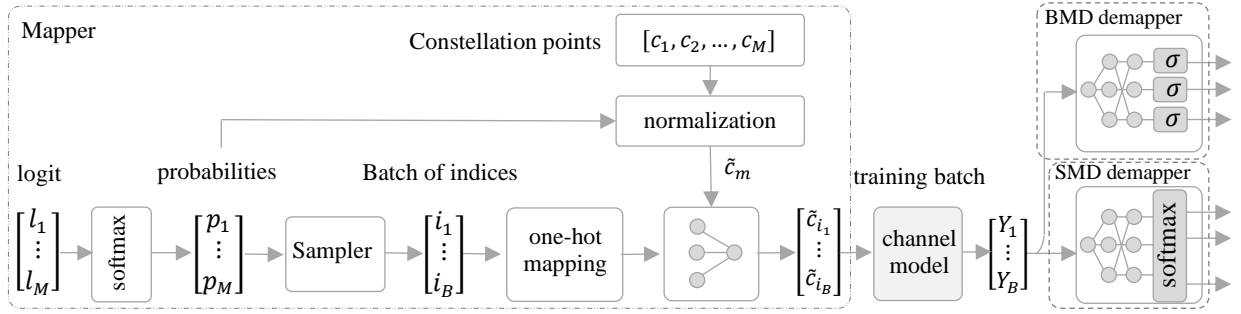
Fig.1: the proposed autoencoder. Two different demappers are considered for SMD or for BMD. $\sigma$ denotes sigmoid activation.

way of symbol mapping is to first encode each index to a one-hot vector of length $M$. The vector is then fed to a single linear layer NN of size $M \times 1$ whose weights are $\tilde{c}_m$. The resulted training batch $[X_1, X_2, \ldots, X_B]$ is sent to the channel. Correspondingly, the channel generates $[Y_1, Y_2, \ldots, Y_B]$ according to some distribution $P(Y|X)$. We assume that a differential channel model is available for the back-propagation of gradients through the channel. Note that as illustrated in Fig. 1, it is more practical to train $l_m = \ln(p_m)$ instead of $p_m$, since $l_m$ is less constraint. However, for simplicity, we will keep using $\mathcal{P}_M$ as trainable parameters in the next derivations. The receiver demapper tries to estimate the transmitted $X_n$ from each received $Y_n$. The two different types of demapper considered are for:

**(i) MI Maximization (SMD):** this demapper is implemented as a categorical classifier by an NN with trainable parameters $\Theta$. It estimates the posterior probability $Q(X_n = c_m | Y_n; \Theta, \mathcal{P}_M, \mathcal{C}_M)$, i.e. how likely $X_n = c_m$ given that $Y_n$ was received, and it is typically trained by minimizing the categorical cross entropy,

$$\mathcal{X}(P, Q; \Theta, \mathcal{P}_M, \mathcal{C}_M) = -\sum_{m=1}^{M} p_m \sum_{Y_n} P(Y_n | c_m) \log_2 \big( Q(c_m | Y_n; \Theta, \mathcal{P}_M, \mathcal{C}_M) \big). \quad (1)$$

Typically, $P(Y_n | c_m)$ is a density function. If this is the case, the inner sum is replaced by an integral over all $Y_n$. As shown in [17,18], the E2E system is trained such to maximize

$$I_{\Theta, \mathcal{P}_M, \mathcal{C}_M} = H(\mathcal{P}_M) - \mathcal{X}(P, Q; \Theta, \mathcal{P}_M, \mathcal{C}_M), \quad (2)$$

where $H(\mathcal{P}_M) = -\sum_{m=1}^{M} p_m \log_2(p_m)$ is the input entropy. Note that $I_{\Theta, \mathcal{P}_M, \mathcal{C}_M}$ is a mismatched MI [19, Eq. 32] and is the achievable rate of the E2E system using SMD. To maximize $I_{\Theta, \mathcal{P}_M, \mathcal{C}_M}$, the autoencoder is trained by computing the derivatives of $I_{\Theta, \mathcal{P}_M, \mathcal{C}_M}$ with respect to (w.r.t.) $\Theta$, $\mathcal{C}_M$ and $\mathcal{P}_M$ via back propagation through all layers of the autoencoder. For derivatives w.r.t. $\Theta$ and $\mathcal{C}_M$, the computation follows the standard minimization of the cross entropy (as in [16-18]). With respect to $p_i$, the derivative of $H(\mathcal{P}_M)$ is $-\log_2(p_i \exp(1))$ and the derivative of $\mathcal{X}(P, Q; \Theta, \mathcal{P}_M, \mathcal{C}_M)$ is

$$\sum_{m=1}^{M} p_m \sum_{Y_n} P(Y_n | c_m) \frac{\partial}{\partial p_i} \big\{ \log_2 \big( Q(c_m | Y_n; \Theta, \mathcal{P}_M, \mathcal{C}_M) \big) \big\} + \sum_{Y_n} P(Y_n | c_i) \log_2 \big( Q(c_i | Y_n; \Theta, \mathcal{P}_M, \mathcal{C}_M) \big), \quad (3)$$

where the first term is computed via back-propagation and then added to the second term. All derivative terms are numerically computed by the empirical averaging over a training batch. Note that the second term of (3) appears since the statistics of input symbols is variable. In categorical classification, the statistics of $M$ categories remain typically the same during training, in which case the second term is not needed. We should emphasize that the second term will be lost if one derives the gradient terms from the sampled (batch) average of $\mathcal{X}(P, Q; \Theta, \mathcal{P}_M, \mathcal{C}_M)$.

**(ii) GMI Maximization (BMD):** GMI is the correct information rate when BMD is applied. Each $c_m$ is mapped to a distinct bit-label $(b_1, b_2, \ldots, b_K)$ with $K = \lceil \log_2(M) \rceil$. This demapper is an NN-based multi-binary classifier with $K$ sigmoid output units and parameters $\Theta$. The output unit of $b_k$ estimates the probability $q(b_k | Y_n; \Theta, \mathcal{P}_M, \mathcal{C}_M)$, i.e. how likely $b_k = 1$ (or 0) given receiving $Y_n$. The autoencoder is trained by maximizing its information rate ([19, Eq. 47])

$$\hat{I}_{\Theta, \mathcal{P}_M, \mathcal{C}_M} = \left[ H(\mathcal{P}_M) + \sum_{k=1}^{K} \sum_{b_k \in \{0,1\}} P(b_k) \sum_{Y_n} P(Y_n | b_k) \log_2 \big( q(b_k | Y_n; \Theta, \mathcal{P}_M, \mathcal{C}_M) \big) \right]^{+}, \quad (4)$$

where $[x]^{+} = \max\{0, x\}$ and $P(b_k)$ is the marginal probability of $b_k$, obtained from $\mathcal{P}_M$ and the bit-labels. Since $P(b_k)$ is variable during the training, the derivative of $\hat{I}_{\Theta, \mathcal{P}_M, \mathcal{C}_M}$ w.r.t. $P(b_k)$ has three terms, like the one of $I_{\Theta, \mathcal{P}_M, \mathcal{C}_M}$ in Eq. 3. Based on these three terms, the batch gradients are computed via back propagation through the autoencoder. Note that in our autoencoder, a fixed bit-labeling is assigned at the beginning of training.

## 3. Simulation Verification and Conclusions

To demonstrate the learning capability of our autoencoder, we verified the learnt PCS and GeoPCS over an additive white Gaussian noise (AWGN) channel. For PCS on the AWGN channels, it is known that the family of Maxwell-

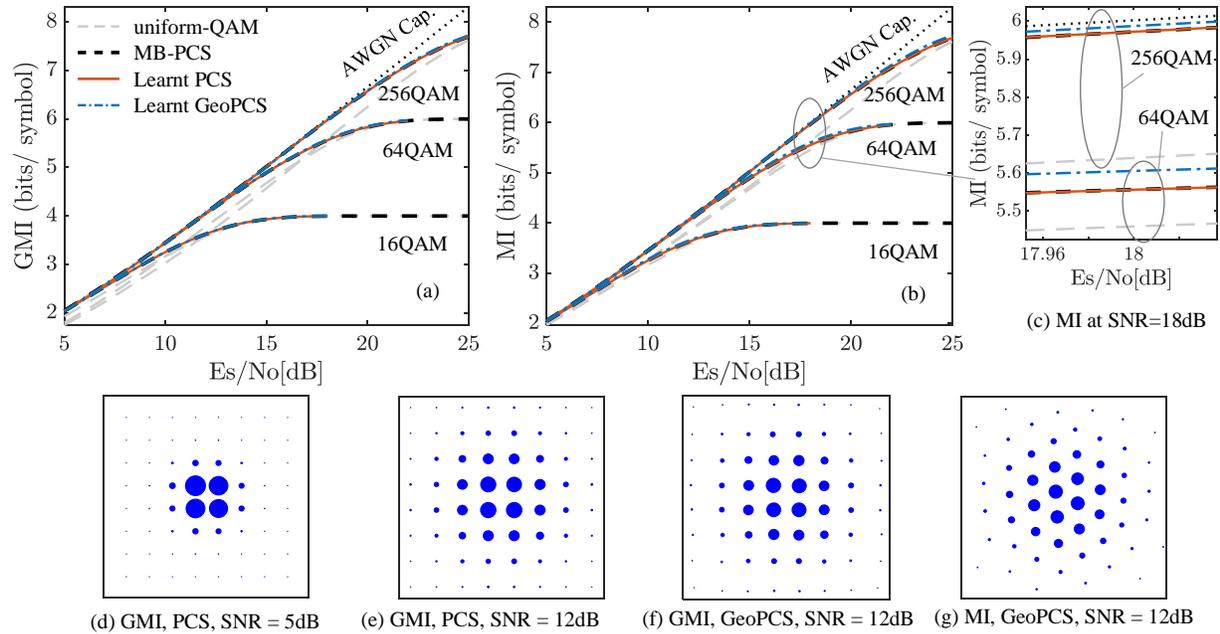

Fig. 2: The learnt PCS and GeoPCS for constellation sizes of M=16, 64, 256 (a) by GMI optimization (b) by MI optimization. The MB-PCS and uniform QAM are shown for comparison. (curves are color coded: dashed black: MB, dashed grey: uniform QAM, solid red: learnt PCS, dash-dotted blue: learnt Geo+PCS). (d-g) The learnt constellation for M=64 at SNR=5dB and 12dB; Points size is proportional to their probability $\mathcal{P}_M$.

Boltzmann (MB) distributions maximizes both the MI and GMI values[19]. Consequently, these theoretical values can be compared with the achieved MI and GMI values after training. For an AWGN channel, it is sufficient to implement the demappers with a rather small NN (see e.g. [16,17]). To speed up the training and enhance the shaping accuracy in our simulations, our demappers used the true posterior distribution $P(X|Y)$ of the AWGN channel (and bit-wise posteriors $P(b_k|Y)$), as explained in [18]. For training, the Adam optimizer was used with a learning rate in the range of $10^{-3} - 10^{-2}$ and the batch size $B$ is chosen between 1000 and 10000. The constellation points ($\mathcal{C}_M$) and their probability ($\mathcal{P}_M$) were initialized by the conventional QAM constellation (the square grid with uniform distribution). For GMI optimization, Gray bit-labeling was used which is optimal for PCS on AWGN channel.

Fig.2 illustrates the learnt PCS (solid red curves) and GeoPCS (dash-dotted blue curves) when GMI (Fig.2(a)) or MI (Fig.2(b)) is maximized. Those curves should be compared with the optimal MB PCS curves (dashed black curves) and the MIs and GMIs of uniform QAM (dashed grey curves). We can see that MIs and GMIs of the learnt PCS are very close to the ones from the theoretical MB-PCS, and the MIs and GMIs of GeoPCS are slightly larger as expected.

The above results show the exactness of our method in directly computing the gradients of the constellation points probabilities with no approximation via a learned autoencoder mimicking a transmission channel. Although we showed here its excellent performance on AWGN channel only, our autoencoder can be applied to different channel models.